# Tuning the Magnetism in Ultrathin $Cr_xTe_y$ Films by Lattice Dimensionality

Guangyao Miao[1#], Minghui Gu[1,2#], Nuoyu Su[1,2], Weiliang Zhong[1,2], Zhihan Zhang[1,2], Yugui Yao[3], Wei Jiang*[3], Meng Meng*[1,4], Weihua Wang*[1,4], Jiandong Guo*[1,2]

[1] Beijing National Laboratory for Condensed Matter Physics and Institute of Physics, Chinese Academy of Sciences, Beijing 100190, China

[2] School of Physical Sciences, University of Chinese Academy of Sciences, Beijing 100049, China

[3] Centre for Quantum Physics, Key laboratory of advanced optoelectronic quantum architecture and measurement (MOE), School of Physics, Beijing Institute of Technology, Beijing, China

[4] Songshan Lake Materials Laboratory, Dongguan 523808, China

* Emails: wjiang@bit.edu.cn; mengm@iphy.ac.cn; weihuawang@iphy.ac.cn; jdguo@iphy.ac.cn.

[#]These authors contribute equally to this paper.

## Abstract

Two-dimensional (2D) magnetic transition metal compounds with atomic thickness exhibit intriguing physics in fundamental research and great potential for device applications. Understanding the correlations between their macroscopic magnetic properties and the dimensionality of microscopic magnetic exchange interactions are valuable for the designing and applications of 2D magnetic crystals. Here, using spin-polarized scanning tunneling microscopy, magnetization and magneto-transport measurements, we identify the zigzag-antiferromagnetism in monolayer $CrTe_2$, incipient ferromagnetism in bilayer $CrTe_2$, and robust ferromagnetism in bilayer $Cr_3Te_4$ films. Our density functional theory calculations unravel that the magnetic ordering in ultrathin $CrTe_2$ is sensitive to the lattice parameters, while robust ferromagnetism with large perpendicular magnetic anisotropy in $Cr_3Te_4$ is stabilized through its anisotropic 3D magnetic exchange interactions.

Intrinsic two-dimensional (2D) magnetic atomic crystals, represented by transition metal compounds (TMC), show great potential in fundamental research and spintronic devices [1-4]. The recent discoveries of ferromagnetism in exfoliated van der Waals (vdW) $CrI_3$ monolayer and $Cr_2Ge_2Te_6$ bilayer [5,6] have inspired the experimental and theoretical explorations, and expanded the landscape of magnetism to 2D systems. In the pursuit of room-temperature 2D magnetic atomic crystals, ultrathin non-vdW materials with 3D lattice of magnetic atoms is an arising class besides the vdW materials [7-9]. Generally, the dimensionality of exchange interaction is directly related to the lattice dimensionality of magnetic atoms. Although some theoretical and experimental works have been devoted to investigate the degree of freedom of spins in 2D system, however, the dependence of magnetism on the dimensionalities of exchange interaction among the spins in atomically thin 2D materials has not yet been addressed systematically.

It is advantageous to compare the microscopic exchange interactions in the same material family with either 2D or 3D bonding network of the same magnetic atoms. Chromium telluride ($Cr_xTe_y$) family hosts abundant material phases with various crystalline structures exhibiting distinct electronic and magnetic properties. Their structures can be categorized into layered structures featuring vdW interlayer interactions, such as $CrTe_3$ and $CrTe_2$, and non-layered structures with three-dimensional (3D) chemical bonding networks, including $Cr_2Te_3$, $Cr_3Te_4$, $Cr_5Te_8$, and CrTe, porviding an ideal system to explore the correlations between the magnetic property and the dimensionality of the exchange interactions among magnetic atoms. In previous studies, high-Curie-temperature ferromagnetism and novel magnetic skyrmions have been discovered in $Cr_xTe_y$ [9-27]. When the thickness of $Cr_xTe_y$ approaches the monolayer (ML) limit, theoretical investigations indicated there are complicated magnetic ground states dependending on the thickness, strain, intercalation, lattice distortions and interlayer coupling strength [28-32]. However, experimental characterizations have revealed various and even controversial magnetic ground states in ML-$CrTe_2$ [13,33,34], and the evolution of magnetism of ultrathin $Cr_xTe_y$ with lattice dimensionality and exchange interaction is yet to be clarified.

Based on phase-selected molecular beam epitaxy (MBE), we fabricate large-scale atomically thin CrTe$_2$ and Cr$_3$Te$_4$ on graphene/SiC(0001) substrates. Combining spin-polarized scanning tunneling microscopy (SP-STM), macroscopic superconducting quantum interference device (SQUID) and magneto-transport measurements, we identify the zigzag antiferromagnetic (AFM) order in ML-CrTe$_2$, incipient ferromagnetism in bilayer (BL) CrTe$_2$, and robust ferromagnetism in BL-Cr$_3$Te$_4$. Combining density functional theory (DFT) calculations, we further reveal that the change of magnetic ordering in vdW CrTe$_2$ upon thickness is driven by the in-plane (IP) strain relaxation, while the robust ferromagnetism with large perpendicular magnetic anisotropy (PMA) of BL-Cr$_3$Te$_4$ originates from the strong anisotropic magnetic exchange interactions within the 3D lattice.

CrTe$_2$ has a layered structure of CdI$_2$-type with space group P$\bar{3}$m1, while Cr$_3$Te$_4$ has a non-layered structure with space group C/2m, as shown in Figs. 1(a) and 1(b). Therefore, the CrTe$_2$ hosts 2D Cr lattices separated by vdW gaps, but the Cr$_3$Te$_4$ has a 3D Cr lattice. By controlling the MBE parameters (see methods in SM [35]), we have successfully fabricated high-quality, mono-phased ultrathin CrTe$_2$ and Cr$_3$Te$_4$ films with different Cr lattice dimensionalities. Figs. 1(c)-(e) show the large-scale STM images of the epitaxial ML-CrTe$_2$, BL-CrTe$_2$, and BL-Cr$_3$Te$_4$ films, indicating that they all follow the 2D growth mode and uniformly spreading over the substrate. The ML-CrTe$_2$ on graphene has a height of 0.98 nm along the $<001>$ direction, while the height of the second-layer CrTe$_2$ is 0.62 nm. The difference originates from the variation of the vdW gaps between first-layer CrTe$_2$/second-layer CrTe$_2$ and ML-CrTe$_2$/graphene, which widely exists in transition metal dichalcogenides grown on graphene/SiC substrate [33,36,37]. The BL-Cr$_3$Te$_4$ has a nominal thickness of 1.60 nm. Note that the growth modes of the initial CrTe$_2$ and Cr$_3$Te$_4$ layers are quite different. The CrTe$_2$ grows in a layer-by-layer mode, but the Cr$_3$Te$_4$ shows a bilayer growth mode at the first two unit cells. Further growth results in a uniform step height of ~0.62 nm in CrTe$_2$ films, and three kinds of steps with heights of ~0.66 nm, 0.35 nm, and 0.31 nm in Cr$_3$Te$_4$ films (See Fig. S1 in SM [35]), consistent with their crystal structures.

As shown in Figs. 1(f)-1(h), the ML-CrTe$_2$ and BL-CrTe$_2$ show similar triangular lattice but different lattice constants ($a$~ 0.37 for ML-CrTe$_2$ and a~ 0.39 nm for BL-CrTe$_2$). The change of IP lattice constant originates from the relaxation of epitaxial strain induced by the graphene/SiC substrate similar to the refer [33] has claimed. The epitaxial strain in first-layer CrTe$_2$ is relaxed in the second-layer due to the presence of vdW gap. In contrast, the BL-Cr$_3$Te$_4$ shows a rectangle unit cell with lattice constants $a$~0.39 nm and $b$~0.69 nm. The surface 2 × 1 modulation of Cr$_3$Te$_4$ arises from the electronic modulation of the periodic Cr atoms between the CrTe$_2$ sub-layers (Fig. 1(b)), and is consistent with previous reports [34,38,39].

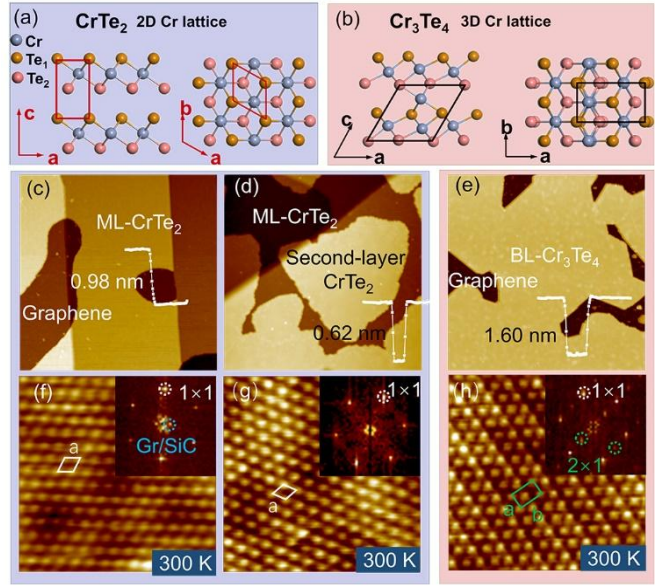

FIG. 1. Phase-selected growth of atomically thin Cr$_x$Te$_y$ films. (a) and (b) Crystal structures of CrTe$_2$ and Cr$_3$Te$_4$. (c)-(e) Large-scale STM images of ML-CrTe$_2$ ($V_b$= -2 V, $I_t$= 50 pA, 140×140 nm$^2$), BL-CrTe$_2$ ($V_b$= -2 V, $I_t$= 50 pA, 100×100 nm$^2$), and BL-Cr$_3$Te$_4$ ($V_b$= -1 V, $I_t$= 50 pA, 180×180 nm$^2$) films. (f)-(h) The atom-resolved STM images of ML-CrTe$_2$ ($V_b$= -2.5 mV, $I_t$= 4 nA ), BL-CrTe$_2$ ($V_b$= 4 mV, $I_t$= 1 nA ), and BL-Cr$_3$Te$_4$ ($V_b$= 3 mV, $I_t$= 5 nA ) films at 300 K. (size of (f)-(h): 5×5 nm$^2$) The insets show their corresponding fast Fourier transform (FFT) images.

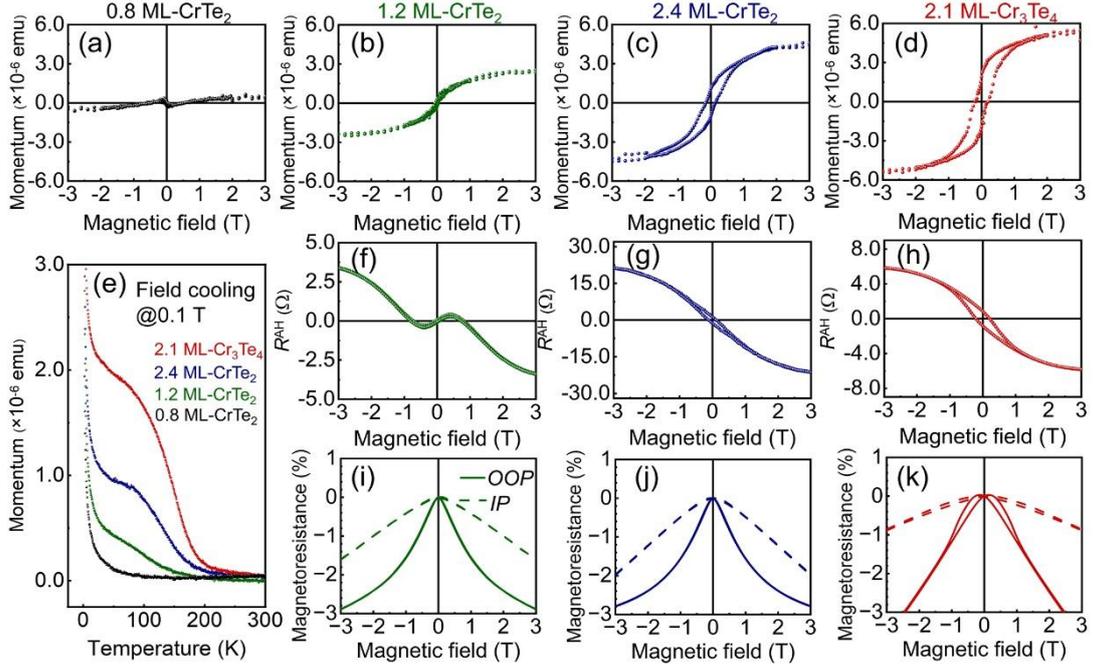

FIG. 2. Magnetic properties of atomically thin $Cr_xTe_y$ films. (a)-(d) Perpendicular magnetic hysteresis loops **M-H** measured at 10 K for $Cr_xTe_y$ films. (e) Temperature dependence of magnetization **M-T** curves for $Cr_xTe_y$ films under field cooling (FC) measured by applying an out-of-plane (OOP) magnetic field **H** = 0.1 T. (f)-(h) Magnetic field dependence of anomalous Hall resistance $R^{AH}$ of 1.2 ML-$CrTe_2$, 2.4 ML-$CrTe_2$, and 2.1 ML-$Cr_3Te_4$ films at 10 K. (i)-(k) Magnetoresistance of 1.2 ML-$CrTe_2$, 2.4 ML-$CrTe_2$, and 2.1 ML-$Cr_3Te_4$ films detected under an in-plane (IP) and OOP magnetic field at 10 K.

The synthesis of large-scale monophase $Cr_xTe_y$ films make it possible to explore the magnetic ground states of the $Cr_xTe_y$ ultrathin films by macroscopic methods including SQUID and magneto-transport. As plotted in Figs. 2(a)-(d), the *MH* curves along out-of-plane (OOP) direction at 10 K for 1.2 ML-$CrTe_2$, 2.4 ML-$CrTe_2$, and 2.1 ML-$Cr_3Te_4$ show typical ferromagnetic (FM) hysteresis character, which is missing in 0.8 ML-$CrTe_2$ (See Fig. S2 in SM for their large-scale STM images) [35]. Their $T_c$ are determined from the *MT* curves (Fig. 2(e)): 153 K for 1.2 ML-$CrTe_2$, 156 K for 2.4 ML-$CrTe_2$, and 183 K for 2.1 ML-$Cr_3Te_4$ (See Fig. S3 in SM for detailed fittings) [35]. For 0.8 ML-$CrTe_2$, no FM signal is observed (black curve in Fig. 2(e)). Note that the rapid increase at low temperature in the four *MT* curves originates from the paramagnetism of graphene [40].

The difference in magnetic properties for atomically thin $Cr_xTe_y$ films are corroborated with complementary magneto-transport measurements. Figs 2 (f)-(h) show their anomalous Hall resistance $R^{AH}$ as a function of the magnetic field at 10 K. The ordinary Hall component $R^{OH}$ has been subtracted as a background (see Fig. S4 in SM for details) [35]. The hysteresis loop is reminiscent of magnetization, indicating the existence of the anomalous Hall effect (AHE) and macroscopic ferromagnetism in 2.4 ML-CrTe$_2$ and 2.1 ML-Cr$_3$Te$_4$ (Figs. 2(g) and 2(h)), which is consistent with the SQUID measurements. For 1.2 ML-CrTe$_2$, there are two additional non-monotonic humps around ±0.5 T and have a large deviation of $R^{AH}$ from the magnetization with a weak hysteresis loop, suggesting the coexisting of AHE and non-trivial hall transport behavior (discussed below). Figs. 2 (i)-(k) compare the magnetic resistance for 1.2 ML-CrTe$_2$, 2.4 ML-CrTe$_2$, and 2.1 ML-Cr$_3$Te$_4$ films under different magnetic fields applied perpendicularly and parallelly to the film surface at 10 K. The magneto-resistances (MR) of all samples along both directions are negative, and the MR under OOP magnetic field for Cr$_3$Te$_4$ shows a butterfly shape characteristic indicating robust ferromagnetism. The value of MR under OOP magnetic field is larger than that under an IP magnetic field at ±3.0 T for each sample, indicating the easy axis of magnetization is along the the OOP direction. As shown in Fig. S2 [35], there are some second-layer CrTe$_2$ islands on the continuous ML-CrTe$_2$ films in the 1.2 ML-CrTe$_2$ sample, while the amount and area of second-layer CrTe$_2$ for the 0.8 ML-CrTe$_2$ sample is negligible, which indicate that the FM signal in 1.2 ML-CrTe$_2$ may originate from the separated second-layer CrTe$_2$ islands (discussed below).

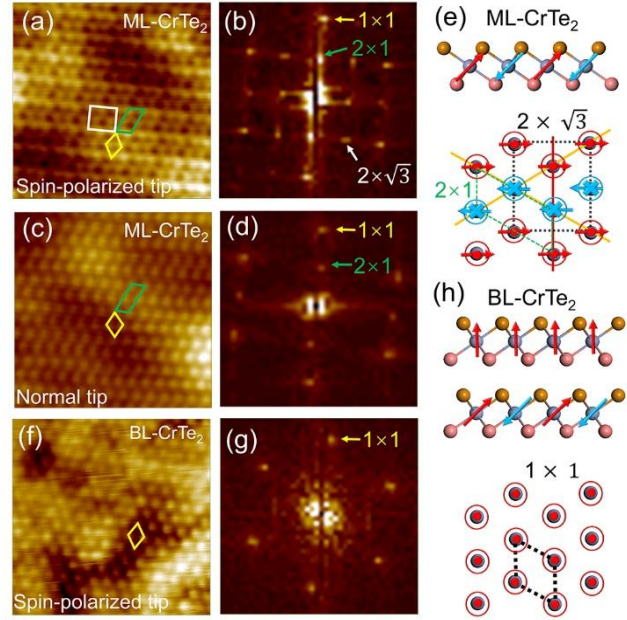

FIG. 3. Spin-polarized STM measurements of the ML-CrTe$_2$ and BL-CrTe$_2$. (a) and (b) The atom-resolved STM image of ML-CrTe$_2$ scanned with a spin-polarized tip ($V_b$= -100 mV, $I_t$= 500 pA, $T$= 5 K, 5×5 nm$^2$) and its FFT image. (c) and (d) The atom-resolved STM image of ML-CrTe$_2$ acquired with a normal Pt/Ir tip ($V_b$= -11 mV, $I_t$= 500 pA, $T$= 78 K, 5×5 nm$^2$) and its FFT image. (e) The spin configuration of ML-CrTe$_2$. The red dot (blue cross) within a circle indicates the spin orientations perpendicular to the layer and pointing outward (inward) to the paper. (f) and (g) The atom-resolved STM image and corresponding FFT image of BL-CrTe$_2$ acquired with a spin-polarized tip ($V_b$= -0.2 mV, $I_t$= 3 nA, $T$= 5 K). (h) The spin configuration of BL-CrTe$_2$.

To explore the magnetic ground state of ML-CrTe$_2$ and to directly compare the magnetic ordering in ML-CrTe$_2$ and BL-CrTe$_2$, we conducted spin-polarized STM experiment. By picking up a Cr cluster on the tip or applying a voltage pulse to drop the cluster, we can reversibly obtain a spin-polarized tip (Cr/Pt/Ir) or a normal tip (Pt/Ir), similar to the previous report [41]. When scanned with a spin-polarized tip, the ML-CrTe$_2$ shows a $2 \times \sqrt{3}$ superstructure at 5 K, as displayed in Figs. 3(a) and (b), in good agreement with the previously reported SP-STM results and the theoretical prediction of an intralayer zigzag AFM ground state [33,42]. (This $2 \times \sqrt{3}$ superstructure persists up to 78 K in our experiments.) When scanned with a normal tip, as shown in Figs. 3(c) and (d), the ML-CrTe$_2$ displays a $2 \times 1$ structral modulation up to 78 K, similar to the image acquired at 5 K in Ref. [33]. There are three equivalent directions with an angle of 120° of the $2 \times 1$ modulation (See Fig. S5 in SI for details [35]). The $2 \times 1$ structural modulation resolved by a normal tip and the $2 \times \sqrt{3}$ superstructure by a spin-polarized tip in ML-CrTe$_2$ at 78 K are in sharp contrast to the triangular $1 \times 1$ lattice at 300 K (Fig. 1(f)), suggesting a structural transition occurs between 300 K and 78 K in ML-CrTe$_2$. As indicated by a recent theoretical work, the ML-CrTe$_2$ hosts strong magneto-elastic coupling phenomenon. Imposing different magnetic states would drive structural transitions with different atomic displacements and deformations of the unit cell can emerge [42]. The observed structural transition in ML-CrTe$_2$ is coupled to its zigzag-AFM state, which may break the equivalence of $a_1$, $a_2$, and $a_3$, and result in two $<\downarrow\uparrow\uparrow\downarrow>$ and one $<\uparrow\downarrow\uparrow\downarrow>$ Cr chains, as indicated by the orange and red lines in the lower panel of Fig. 3(e). The spin-spin interactions accompany with strong magneto-elastic coupling would drive the dimerization of the Cr atoms along the two $<\downarrow\uparrow\uparrow\downarrow>$ lines, but keeping the initial lattice constant along the $<\uparrow\downarrow\uparrow\downarrow>$ line, leading to a $2 \times 1$ superstructure.

However, the $2 \times 1$ and $2 \times \sqrt{3}$ superstructures of ML-CrTe$_2$ are missing in the second-layer CrTe$_2$. The atom-resolved STM image acquired on the second-layer CrTe$_2$ using a spin-polarized tip shows a triangular $1 \times 1$ structure as displayed in Figs. 3(f) and (g). Combining the SP-STM, SQUID, and magneto-transport results, BL-layer CrTe$_2$ has a FM ground state with OOP magnetic easy-axis, which is missing in ML-CrTe$_2$. The FM signal of BL-CrTe$_2$ sample originates from the second-layer CrTe$_2$. Its FM magnetic configuration is schematically illustrated in Fig. 3(h). The zigzag AFM magnetic ground state of ML-CrTe$_2$ is revealed by SP-STM, without any FM signal detected in the 0.8 ML-CrTe$_2$ sample. The observed weak FM signal in the 1.2 ML-CrTe$_2$ sample originates from the second-layer CrTe$_2$ islands with FM ground state. Furthermore, the lateral edges (vertical vdW gap) between the second-layer CrTe$_2$ and ML-CrTe$_2$ are the boundary (interface) of FM and AFM domains, together with those orientational domains walls shown in Fig. S5 [35], could host topological spin texture and induce nontrivial transport behavior, as suggested by the magneto-transport data of 1.2-ML CrTe$_2$ (Fig. 2(f)). The two hump features around ±0.5 T may originate from the interactions between the transport carriers and the potential real-space chiral spin-textures [43].

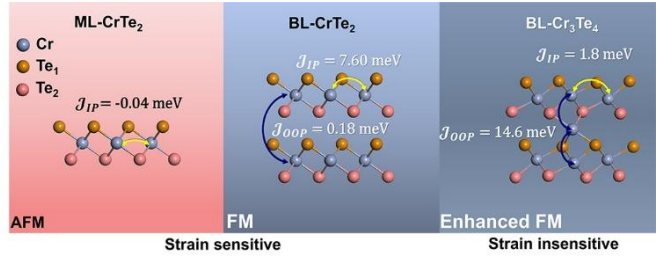

FIG. 4. The evolution of the magnetic ground states and magnetic exchange coupling in atomically thin Cr$_x$Te$_y$ systems, calculated with the following lattice constants: $a$=0.370 nm for ML-CrTe$_2$, $a$=0.385 nm and $c$=0.620 nm for BL-CrTe$_2$, $a$=0.396 nm, $b$= 0.685 nm, and $c$=0.658 nm for BL-Cr$_3$Te$_4$.

Our experimental results evidence that the magnetic properties of Cr$_x$Te$_y$ are related to their lattice dimensionality. To reveal the microscopic mechanism, we carried out DFT calculations to investigate the magnetic states of Cr$_x$Te$_y$ in different phases with varied lattice parameters. Simplified Heisenberg model considering the nearest neighbor term $\mathcal{H} = \sum_{<i,j>} \mathcal{J}_{ex} S_i S_j$ is used to describe the localized magnetic moment and mapped with the DFT-calculated total energy, based on which the magnetic exchange interaction constant $\mathcal{J}_{ex}$ can be determined ($\mathcal{J}_{ex} > 0$ and $\mathcal{J}_{ex} < 0$ correspond to the FM and AFM coupling, respectively) [44].

First, we studied the 2D system ML-CrTe$_2$ with only the IP intralayer exchange interaction ($\mathcal{J}_{IP}$). We find that the $\mathcal{J}_{IP}$ sensitively depends on the IP lattice constant [31,45]: When the lattice parameters of the bulk CrTe$_2$ ($a$ = 0.377 nm) are adopted [11], a positive $\mathcal{J}_{IP}$ (FM state) is obtained for ML-CrTe$_2$. Decreasing the IP lattice parameters by less than 1% results in a sign change of $\mathcal{J}_{IP}$, indicating that the transition from FM to AFM is induced by compressive strain (See Fig. S6 in SI for details) [35]. With the IP lattice parameters exceeds the bulk values, the positive $\mathcal{J}_{IP}$ further increases, and results in a stable FM state. Thus the experimentally observed AFM in ML-CrTe$_2$ in the current work is ascribed to the slightly compressive strain from the graphene/SiC substrate. Our calculations also explain the different results of the magnetic ground states of ML-CrTe$_2$ by different research groups — the different substates used for ML-CrTe$_2$ grown may give different IP strain [13, 33,34].

Second, we evaluated the magnetic states of the BL-CrTe$_2$, in which both of the IP exchange interaction ($\mathcal{J}_{IP}$) and the OOP exchange interaction ($\mathcal{J}_{OOP}$) are considered. Note $\mathcal{J}_{OOP}$ describes the magnetic interaction between Cr atoms mediated by Te atoms in neighboring layers, and therefore is sensitive to the interlayer distance. Indeed the sign of $\mathcal{J}_{OOP}$ does reverse from negative to positive when the interlayer distance increases from 0.598 nm to 0.620 nm (see Fig. S7 in SM for details) [35]. More importantly, the change of $\mathcal{J}_{IP}$ upon the increase of the IP lattice parameters is more significant than that of $\mathcal{J}_{OOP}$ upon the increase of the interlayer distance. Considering the experimentally observed increase of the IP lattice parameters of the second layer CrTe$_2$ ($a$~0.39 nm) relative to the first layer ($a$~0.370 nm), the FM of the BL-CrTe$_2$ should be mainly attributed to the strain relaxation effect. The machanism for the observed FM magnetic states is schematically shown in the left and middle panels of Fig. 4.

We further studied the Cr$_3$Te$_4$ system with 3D Cr lattice in both bulk and bilayer thin film forms. The distinct 3D magnetic interaction in Cr$_3$Te$_4$ is fundamentally different from that in 2D vdW CrTe$_2$ systems. The results show robust FM ground state with large $\mathcal{J}_{OOP}$ and $\mathcal{J}_{IP}$ independent of lattice parameters, which is attributed to the strong FM exchange coupling in the 3D Cr network as illustrated in the right panel of Fig. 4. Similar to the magnetic ground states, the MA of BL-Cr$_3$Te$_4$ is quite robust against lattice variations, and the calculated PMA energy is as high as 21.5 meV. In constrast, the MA of CrTe$_2$ systems exhibits high sensitivity to the lattice parameters.

In summary, we achieved the high-precision control of the molecular beam epitaxy of ultrathin CrTe$_2$ and Cr$_3$Te$_4$ films. The microscopic SP-STM and macroscopic measurements SQUID and magneto-transport measurements evidence the zigzag antiferromagnetism in the ML-CrTe$_2$, ferromagnetism introduced by second-layer CrTe$_2$ in the BL-CrTe$_2$, and robust ferromagnetism with large PMA in BL-Cr$_3$Te$_4$. We further revealed that the magnetic ground states of atomically thin

CrTe$_2$ are sensitive to the lattice constants (strain) and electronic correlations by DFT calculations, demonstrating its advantages in the manipulations of magnetic states. The nontrivial hall transport in 1.2 ML-CrTe$_2$ implyies that this system could host topological spin-texture with potential applications in spintronic devices. As for the atomically thin Cr$_3$Te$_4$, the strong anisotropic 3D exchange interactions result in robust ferromagnetism and PMA against the lattice strain and external perturbation. Our research emphasizes the role of the dimensionality of the exchange interactions in tuning the magnetism of magnetic transition metal compounds, and sheds light on the design of magnetism in ultrathin films with the thickness toward 2D limits.

## Acknowledgements

This work was supported by the National Natural Science Foundation of China (Grants No. 11974399, 11974402, 12104494, and 12061131002), the Strategic Priority Research Program of Chinese Academy of Sciences (Grants No. XDB33000000 and XDB30000000), and the National Key R&D Program of China (2022YFA1403000, and 2021YFA0718700). W.J. acknowledges the support from NSF of China (Grant No. 12204037) and the Beijing Institute of Technology Research Fund Program for Young Scholars., M.M. was supported by the Youth Innovation Promotion Association, CAS. G.M. acknowleges the support from the IOP-Humboldt postdoc fellowship in physics of institute of physics, CAS and Integrative Research Institute for the Sciences of Humboldt-Universität zu Berlin.